\documentclass[12pt, english, titlepage]{article}

\usepackage[english]{babel}
\usepackage{graphicx}
\usepackage[pdfborder={0 0 0}]{hyperref}
\usepackage[font = small, labelfont = bf]{caption}
\usepackage[square, comma]{natbib}
\setcitestyle{numbers}
\RequirePackage{doi}
\usepackage{hyperref}

\usepackage{verbatim}

\usepackage{amsmath}
\usepackage{amssymb}  
\usepackage{amsthm}     
\usepackage{float}

\usepackage{xcolor}

\usepackage[affil-it]{authblk}

\begin{document}

\title{Diversity and its decomposition into variety, balance and disparity}
\author{Alje van Dam\thanks{Email: \texttt{A.vanDam@uu.nl}}}
\affil{Copernicus Institute for Sustainable Development, Utrecht University\\
Centre for Complex Systems Studies (CCSS), Utrecht University}

\maketitle

\begin{abstract}
Diversity is a central concept in many fields. Despite its importance, there is no unified methodological framework to measure diversity and its three components of variety, balance and disparity. Current approaches take into account disparity of the types by considering their pairwise similarities. Pairwise similarities between types do not adequately capture total disparity, since they fail to take into account in which way pairs are similar. Hence, pairwise similarities do not discriminate between similarity of types in terms of the same feature and similarity of types in terms of different features. This paper presents an alternative approach which is based similarities of features between types over the whole set. The proposed measure of diversity properly takes into account the aspects of variety, balance and disparity, and without having to set an arbitrary weight for each aspect of diversity. Based on this measure, the 'ABC decomposition' is introduced, which provides separate measures for the variety, balance and disparity, allowing them to enter analysis separately. The method is illustrated by analyzing the industrial diversity from 1850 to present while taking into account the overlap in occupations they employ. Finally, the framework is extended to take into account disparity considering multiple features, providing a helpful tool in analysis of high-dimensional data. 

\end{abstract}

\section{Introduction} 
Diversity is a central concept in a wide range of scientific fields. In the natural sciences, it is often associated with the functional properties of a system, like the stability of ecosystems \cite{MacArthur1955, Tilman2014}. In the social sciences, the concept of diversity is key to theories regarding recombinant innovation \cite{VandenBergh2008, Weitzman1998}, regional development \cite{Frenken2007}, cultural evolution \cite{Foley2011}, and the science of science \cite{Rafols2010, Zang2016}.

But what exactly is diversity and how can it be measured? Recent frameworks emphasize that diversity consists of three dimensions \cite{Daly2018, Page2011, Purvis2000, Stirling2007}. First, the \textit{variety} describes the number of different types, species or categories present.\footnote{I follow the terminology used in \cite{Stirling2007}, but these concepts are known by different names in different fields, for example as 'richness', 'evenness' and 'similarity' in ecology.} The variety is bounded by the total number of types in the classification or taxonomy that is used. Second, the \textit{balance} describes how individuals or elements are distributed across these types. When elements are concentrated in few types the balance is low, whilst a high balance indicates a more even distribution. Last, the \textit{disparity} takes into account to what extent the types considered differ from each other in terms of some given features or characteristics. If the types considered are very similar, they have low disparity. An increase along any of these three dimensions corresponds to an increase in overall diversity. A proper measure of diversity should therefore take into account all three dimensions. 

Despite the importance of diversity as a concept, there is no unified methodological framework to measure and analyze the three dimensions of diversity. In the past, the disparity was not even considered by most diversity indices. More recently, there have been attempts to incorporate disparity into a measure of diversity by including some measure of the pairwise distances or similarities between the types considered. An example is Rao's quadratic entropy \cite{Rao1982}, introduced into the social sciences in \cite{Stirling2007} where it is known as the Rao-Stirling diversity. It expresses diversity as the average distance between types weighed by their relative frequencies. Recently, it has been shown that Rao's quadratic entropy can be derived from a more general class of diversity measures  that take into account pairwise similarities between the types \cite{Leinster2012}.

However, using pairwise similarities to measure diversity leads to both practical and conceptual problems. One practical problem is that there are many different ways in which pairwise similarities can be inferred from given data \cite{Eck2009, Yildirim2014}, so any diversity measure based on pairwise similarities is subject to an ad-hoc choice of a particular similarity measure. In addition, it is unclear how heavily such an index should weigh disparity versus variety and balance  \cite{Stirling2007}. Most importantly, considering only pairwise similarities between types may not adequately capture total disparity, since pairwise similarities do not take into account \textit{in which way} pairs are similar. Using pairwise similarities, there is no way to distinguish situations in which all types are similar \textit{in terms of the same feature} as opposed to them being similar in terms of different features. Both could lead to a different value of the diversity. 

This paper presents a framework to measure diversity that does not rely on pairwise similarities between types. Instead, disparity is taken into account by looking at the overlap of features between types over the whole set. This is done by drawing on the concepts of alpha, beta and gamma diversity from ecology \cite{Whittaker1972} and the corresponding decomposition of diversity as introduced in \cite{Jost2007}, which is based on Hill numbers \cite{Hill1973}. The result is a measure of diversity that incorporates variety, balance and disparity simultaneously, and has a natural interpretation as the 'number of compositional units' \cite{Tuomisto2010}. 

Building on this measure, I introduce the 'ABC decomposition' that allows one to decompose the diversity into separate measures of variety, balance and disparity. This enables the study of the distinct role each of these dimensions has in different systems. The proposed framework is closely related to information-theoretic measures of uncertainty, and the use of multivariate information theory shows how the measure can be extended to take into account disparity along multiple dimensions. This leads to two results regarding the diversity of types given multiple feature sets, depending on the dependence structure of the variables involved. First, diversity considering multiple features becomes multiplicative when different features are independent. Second, additional features may be neglected in measuring diversity in the case that the second feature set and the types are conditoinally independent given the first set of features. 

I proceed as follows. Section \ref{sec:decomp} starts with an example of a situation where using pairwise similarities fails to quantify disparity correctly. Subsequently the concepts of beta diversity are introduced along with the main result, namely a measure of diversity that takes into disparity as the overlap over a set of features. Section \ref{sec:ABC} then introduces a decomposition of diversity into seperate measures of variety, balance and disparity. As an illustration, I apply the proposed measures to historical data in order to characterize the change in diversity of industries in the US, taking into account disparity in terms of the occupations that industries employ. Section \ref{sec:infotheory} shows how the framework can be extended to take into account multiple sets of features. I conclude with a brief discussion of the results. 

\section{Decomposing diversity} \label{sec:decomp}

\paragraph{An example}\label{sec:example}
Consider a region in which certain economic activities takes place in the form of industries. These industries can be thought of to consist of a certain set of inputs or features \cite{Hidalgo2009}. We will represent these features with letters in a set $S$, and the industries as words in set $S'$. For example, one might think of the letters as occupations required by a firm to engage in a particular industry, represented as a word. The diversity of words is determined by the number of different words (variety), their relative frequency (balance) and their similarity in terms of the letters they consist of (disparity). Adding words with similar composition of letters does not affect the diversity much, whereas adding words consisting of many new letters may greatly increase diversity. 

\begin{figure}[h]
\includegraphics[width=.5\linewidth]{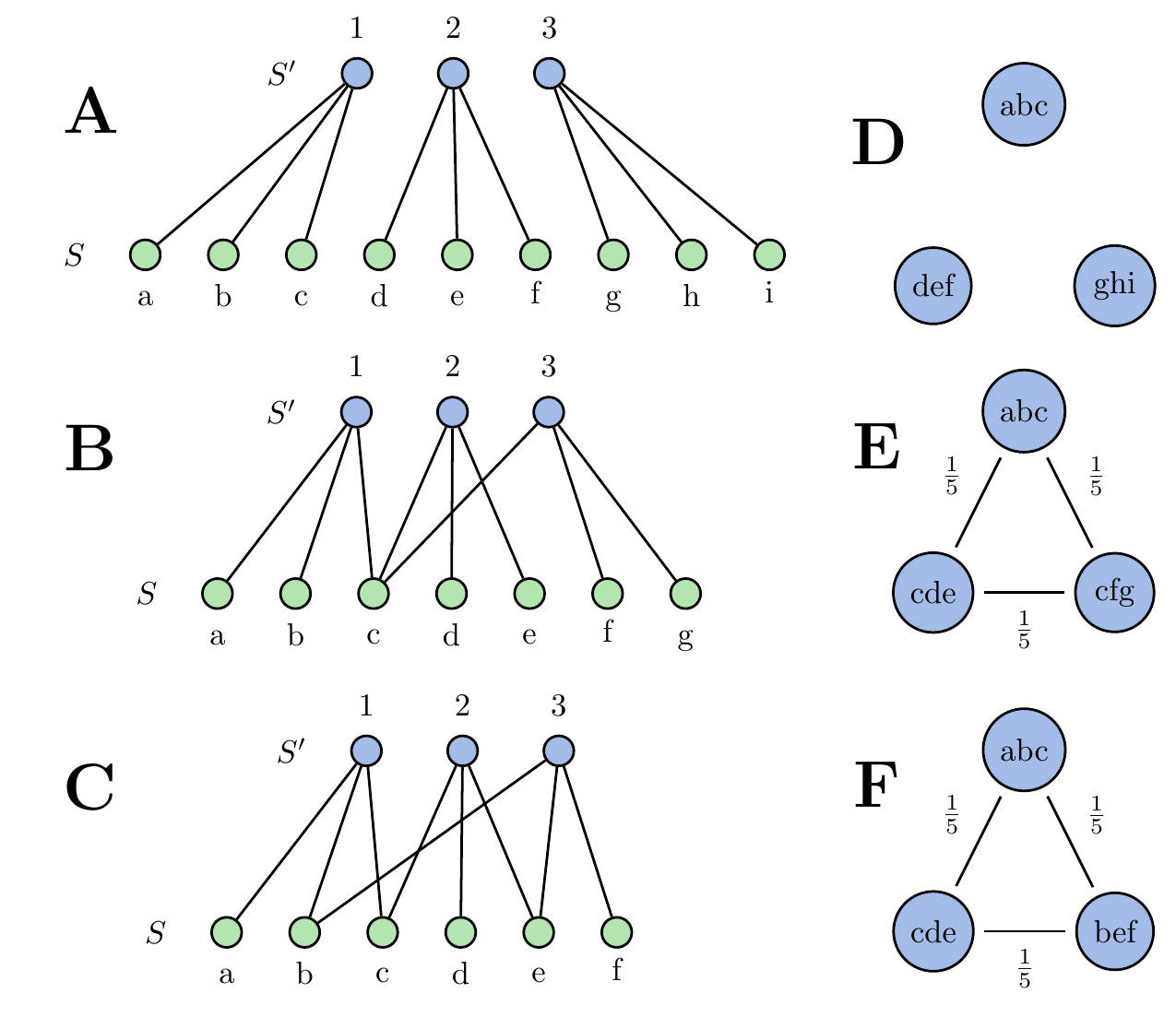}
\caption{{\bf A}, {\bf B} and {\bf C} show the bipartite networks as discussed in the main text. One can think of the blue nodes respresenting three industries (words), and the green nodes representing nine occupations (letters) that characterize the industries. {\bf D}, {\bf E} and {\bf F} show the corresponding projected industry networks, in which the edge weights are given by the Jaccard similarity between industries. In {\bf A} and {\bf D} there is no overlap in occupations, and pairwise similarities between industries are $0$, as shown by the absence of edges in {\bf D}. The Rao-Stirling diversity is given by $\Delta = \frac{1}{3}$. {\bf B} and {\bf E} show a situation where the industries use a total of seven occupations, and the similarity between each industry equals $s_{ij} = \frac{1}{5}$. {\bf C} and {\bf F} show a situation where only six occupations are present, and where all pairwise similarities are again $\frac{1}{5}$. Although {\bf B} has one occupation more than {\bf C}, their projections {\bf E} and {\bf F} are identical, and therefore any diversity measure based on pairwise similarities will assign identical diversities to both cases. The Rao-Stirling diversity is given by $\Delta = \frac{4}{15}$ for both cases.}

\label{fig:example}
\end{figure}

The composition of words and letters in a region can be represented as a bipartite network as in Figure \ref{fig:example}.
In the three cases shown, the variety equals $3$ (there are three unique words) and the balance is maximal (the relative frequency $p_i = \frac{1}{3}$ for each word). The disparity of words is different for each of the three cases, and is determined by how the words are composed from the letters. 

A common approach to quantify diversity whilst taking into account disparity is by considering the pairwise similarity between types \cite{Leinster2012, Rao1982, Stirling2007}. Computing the pairwise similarities can be interpreted as 'projecting' the bipartite network onto a weighted network in which the nodes are the types, and the wighted edges represent the pairwise similarities in terms of the overlap in features (see Figure \ref{fig:example}). Here we consider the Jaccard similarity $s_{ij}$, which gives the similarity as the numer of shared features divided by the total number of features used by both types. 

An example of such a measure is the Rao-Stirling diversity, which is computed as\footnote{$1-s_{ij}$ gives the 'Jaccard distance' or dissimilarity between a pair of words.} \cite{Rao1982, Stirling2007}
\begin{align*}
\Delta = \sum_{ij} (1 - s_{ij}) p_i p_j.
\end{align*}
This measure incorporates the variety by summing over all types, and the balance by taking into account the relative frequencies $p_i$. Disparity is then taken into account by weighing every pair of types by the distance between the types. This way, pairs with low similarity contribute more to the diversity than pairs with high similarity. 

In the first case in Figure \ref{fig:example} the disparity is maximal (there is no overlap of letters between words), and the Rao-Stirling diversity reduces to $\Delta = \sum_{ij} \frac{1}{3}\frac{1}{3} = \frac{1}{3}$ since $s_{ij}=0$ for all pairs. For the other two cases, the Jaccard similarities are given by $s_{ij} = \frac{1}{5}$ for all pairs. Since the pairwise similarities are identical in both cases, any diversity measure based on these pairwise similairities will give the same diversity for both cases. Indeed, computation of Rao-Stirling diversity shows a diversity of $\Delta = \sum_{ij} (1-\frac{1}{5})\frac{1}{3}\frac{1}{3} = \frac{4}{15}$ for both cases. 

However, note that the number of features in the latter two cases in Figure \ref{fig:example} is different. Out of two collections of features with similar variety and balance, the collection that consists of more different features is arguably more diverse. Hence, we expect a higher diversity in middle case in Figure \ref{fig:example}. Since the projected networks for the middle and last case are identical, this difference cannot be captured by a diversity measures that is based on those similarities. Instead, a diversity measure should take into account the overlap in features over the whole collection of types, as opposed to all pairs. The mutual information-based measure proposed in the current paper accurately reflects the difference in composition between the two cases. 

\paragraph{Hill numbers}
In our measurement of diversity, we build on the framework of Hill numbers, which provides a unifying mathematical framework for the measurement of diversity when disparity is not taken into account \cite{Hill1973, Jost2006}. Hill numbers define diversity as the inverse of a generalized weighted average of the relative frequencies of the types. In this definition, a collection is diverse if the types are on average rare, i.e. the average share of the types is low. 

Hill numbers satisfy a number of axiomatic requirements for a measure of diversity, including symmetry, continuity, and monotonicity in the number of species \cite{Daly2018}. Another key property is the \textit{replication principle}, which states that pooling together two collections that do not share any types but have equal distributions, should give a new collection with double the diversity of the original collections \cite{Hill1973}. 

Hill numbers give rise to a parametric family of diversity measures, in which a parameter $q$ determines how heavily one weighs the rarity of types in a measure of diversity. For $q=1$, rare and common species are weighed equally heavy and the Hill number equals the exponential of the Shannon entropy:
\begin{align} \label{eq:expshan}
D(S) = e^{H(X)} = e^{-\sum_i p_i \log{p_i}}.
\end{align}
Here, $S$ is a collection of elements with types $i$ and relative frequencies $p_i$, and $X$ is a random variable that represents the type $i$ of a randomly drawn element from $S$. A more elaborate discussion on Hill number and their relation to Shannon entropy can be found in Appendix \ref{sec:app_hill}. It was shown in \cite{Jost2007} that the Hill number with $q=1$, i.e. the exponential of the Shannon entropy, is the unique measure that satisfies all axiomatic requirements and allows for a decomposition of independent within- and between components in the presence of groups. 

Hill numbers have also been referred to as the 'true diversity' as opposed to an index, as many existing diversity indices in ecology and economics that were originally introduced based on heuristics have been shown to be a transformation of a Hill number \cite{Jost2006}. In particular, equation \eqref{eq:expshan} shows how the Shannon entropy, a popular \textit{index} of diversity but which is actually a measure of uncertainty (it has units in 'bits' or 'nats'), can be transformed into a measure of diversity \cite{Jost2006}.

Furthermore, the Hill number of a collection has a clear interpretation as the 'effective number' of types, meaning that the Hill number of a collection $S$ can be interpreted as the number of types that would be present in a virtual collection $\tilde{S}$ that has maximal balance (i.e. a uniform distribution over types) and has the same diversity as $S$. In particular, for a uniform dsitribution, i.e. $p_i= \frac{1}{n}$ for all $i$, we have $D(S) = n$ so that the diversity equals the number of types. For any other distribution over types, the Hill number represents the equivalent number of types in a maximally balanced collection. 

\paragraph{Alpha and beta diversity and the number of compositional units}
The Hill number $D(S)$ quantifies both the variety and balance of types but not their disparity, and thus implicitely assumes that all types $i$ are maximally disparate. Here, we aim to extend this framework to include the overlap of features between types. To this end, we build on the concepts of alpha, beta and gamma diversity from ecology. 

Hill numbers provide a decomposition of diversity diversity into its $\alpha$ and $\beta$ components \cite{Jost2007}, which are used in ecology to descibe the average within-sample diversity and the between-sample diversity, respectively \cite{Whittaker1972}. For example, consider a forest in which the distribution of species is sampled in different plots. The diversity of the collection of species that consists of all plots pooled together is called the total diversity or $\gamma$ diversity. The $\alpha$ diversity represents the average diversity \textit{within} each plot. The $\beta$ diversity represents the diversity \textit{between} each plot, reflecting the diversity that is the results of the differences in species composition between each plot.  

The $\gamma$ diversity of the forest can be multiplicatively decomposed into independent $\alpha$ and $\beta$ components, i.e.  $D_{\gamma} = D_{\alpha}D_{\beta}$ \cite{Jost2007}. In a homogenous forest, where all plots have approximately the same species composition, the average within-plot diversity $D_{\alpha}$ is close to the diversity of all plots pooled together, $D_{\gamma}$. Hence, the between-plot diversity $D_{\beta}$ will be close to $1$. In a heterogeneous forest on the other hand, every plot has very different species compositions and contains only a small part of the total diversity, so $D_{\alpha}$ is much smaller than $D_{\gamma}$, leading to a higher value of $D_{\beta}$. 

$D_{\beta}$ reflects the number of different plots needed, each with diversity $D_{\alpha}$, to obtain a pooled diversty of $D_{\gamma}$. The maximum value of $D_{\beta}$ is given by $D_{\gamma}$, corresponding to the case where every plot consists of a unique species ($D_{\alpha} = 1$). The $\beta$ diversity is thus bound from below by $1$ and from above by $D_{\gamma}$. 

Note that the situation described above corresponds with the example in Figure \ref{fig:example}, in which the plots represent the types of interest (words) and the species represent some characterizing features of those types (letters). The values of each diversity for the example are given in Figure \ref{fig:alphabeta}. In this setting, the $\gamma$ diversity gives the total diversity of features, and the $\alpha$ diversity the average diversity of features within a type. The $\beta$ diversity then represents the 'between-type' diversity based on the heterogeneity of the composition of types. It gives the number of types of diversity $D_{\alpha}$ that are needed to obtain a total diversity of features $D_{\gamma}$ when there is no overlap of features between the types. Hence, it can be interpreted as a measure of the 'number of compositional units', giving the effective number of types that would be present when the types do not share any features and would be equally abundant \cite{Tuomisto2010}. Framing $D_{\beta}$ in terms of types and features, it provides a measure of diversity of types that takes into account variety, balance \textit{and} disparity as given by the overlap of features between types.   

\begin{figure}[h] 
\centering
\begin{tabular}{| l | c | c | c | c |}
    \hline
	& $\alpha$-diversity & $\beta$-diversity & $\gamma$-diversity & eff. number\\ 
	& $D_{\alpha}(S)$ & $D_{\beta}(S')$ & $D_{\gamma}(S)$ & $D(S')$ \\ \hline
	$\bf{A}$ & 3& $3$ & $9$ & $3$  \\ 	 \hline
	$\bf{B}$ & 3& $2.08$ & $6.24$ & $3$   \\ 	 \hline
	$\bf{C}$  & 3& $1.89$ & $5.67$ & $3$  \\	 \hline
\end{tabular}

\caption{Values of the $\alpha$, $\beta$ and $\gamma$ diversities for the three examples depicted in Figure \ref{fig:example}. The average diversity of occupations within an industry, $D_{\alpha}(S')$ is equal in all three examples, as every industry employs three different occupations with equal weight and all industries have an equal share. The total diversity of features $D_{\gamma}(S)$ is given by the effective number of occupations in all industries pooled together, and differs in each case. This also leads to different values of $D_{\beta}(S')$. For completeness, the effective number of industries $D(S')$, representing the diversity of industries when one assumes that they are totally disparate, is also included. }
\label{fig:alphabeta}
\end{figure}

\paragraph{Measuring diversity of industries} 
Here, the general application of the proposed diversity measure is presented using an empirical example. The aim is to quantify industrial diversity in the US, where the distinguishing features of industries is considered to be the different occupations they employ. US census data was extracted from IPUMS-USA \cite{Ruggles2018}, providing a 1\% sample\footnote{For $1980$ and $1990$ a 5\% sample was given. Furthermore, note that the analysis presented is for illustrative purposes so further data cleaning and consistency issues are not considered here.} of total population in the US for every decade from $1850$ to $2010$. The data contains for every person their occupation $i \in S$ and industry $j \in S'$. The used classifications consist of $269$ occupation types and $147$ industry types. 

The data is interpreted as a weighted bipartite network as in Figure \ref{fig:example}, with nodes $i$ in the occupation layer $S$ and nodes $j$ in the industry layer $S'$. The edge weight between nodes $i$ and $j$ is given by the number of people $q_{ij}$ working in occupation $i$ and industry $j$. The strength of node $i$ is given by $q_i = \sum_j q_{ij}$ and represents the total employment in occupation $i$, and similarly $q_j$ denotes total employment in industry $j$. Normalizing the quantities $q_i$, $q_j$ and $q_{ij}$ by the total number of people $Q = \sum_{ij}q_{ij}$ gives the relative frequencies $p_{ij} = \frac{q_{ij}}{Q}$, $p_i = \frac{q_i}{Q}$ and $p_j = \frac{q_j}{Q}$ respectively. Each of the relative frequencies may in turn be interpreted as the probability distribution of a random variable that represents the occupation or industry type of a randomly sampled person, i.e. $p_i = P(X=i)$, $p_j= P(Y = j)$ and $p_{ij} = P(X=i, Y=j)$.

Using Hill numbers, the effective number of industries and occupations can be expressed as $D(S') = e^{H(Y)}$ and $D(S) = e^{H(X)}$, respectively. To obtain the effective number of occupations \textit{within} an industry, consider the relative frequencies $p_{i|j} = \frac{q_{ij}}{q_j}$ of occupation $i$ in industry $j$. The occupational diversity of an industry $j$ is then given by
\begin{align*} 
D(S_j) = e^{H(X|j)} = e^{\sum_i p_{i|j} \log(p_{i|j})}.
\end{align*} 
The \textit{average} within-industry diversity is then given by \cite{Jost2007}
\begin{align*}
D_{\alpha}(S) = e^{H(X|Y)} = e^{-\sum_j p_j \sum_i p_{i|j} \log(p_{i|j})},
\end{align*} 
where $H(X|Y)$ is the conditional entropy of $X$ given $Y$. Finally, the \textit{within} industry diversity $D_{\beta}$ follows from multiplicatively decomposing the total occupational diversity $D_{\gamma}(S)$ into its $\alpha$ and $\beta$ components, leading to \cite{Jost2007}
\begin{align*} 
D_{\beta}(S') = \frac{D_{\gamma}(S)}{D_{\alpha}(S)}.  
\end{align*} 
$D_{\beta}(S')$ can be interpreted as the effective number of industries, \textit{discounted for the overlap in their occupational distributions}. Its units correspond to the number of industries that would be present in the case of equally-distributed, non-overlapping industries, and where the $D_{\alpha}$ and $D_{\gamma}$ are the same. 

\begin{figure}[h]
\centering
\includegraphics[width=.8\textwidth]{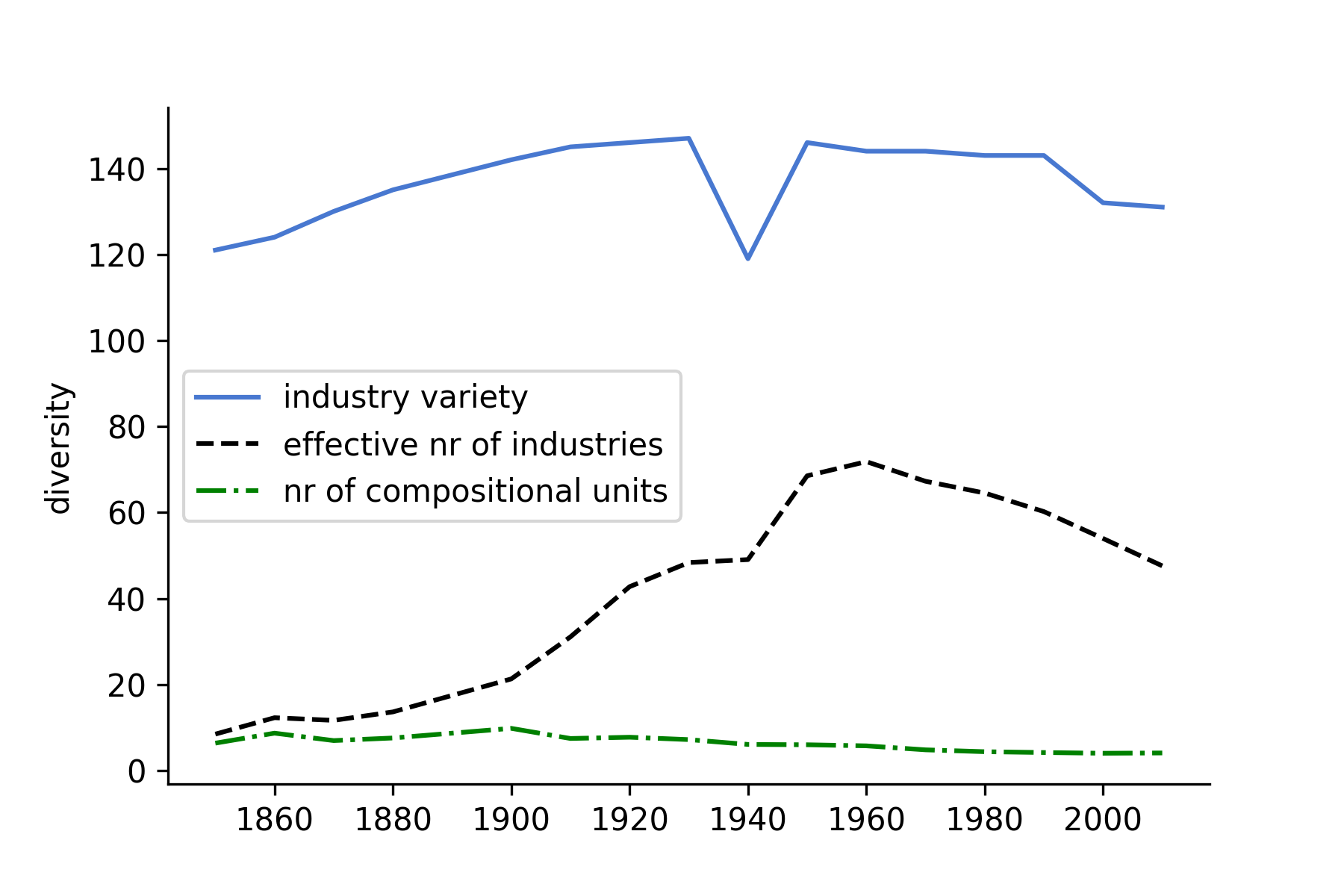} 

\caption{Variety, effective number and effective number of compositional units for industries. The variety of industries is approximately constant over time. The effective number of industries takes into account variety and balance and shows a hump-shaped pattern, where initially the distribution of people over industries becomes more equal reaching a diversity of $80$ effective industries in 1960, where a reconcentration starts to take place. The diversity takes into account the occupational overlap between industries in terms of compositional units. In 1850 the the industrial diversity was equivalent to approximately ten non-overlapping industries, which declined to approximately four compositional units in 2000. }
\label{fig:IPUMS_divs}
\end{figure}

Figure \ref{fig:IPUMS_divs} shows the time evolution of variety, the effective number of industries $D(S')$ (taking into account variety and balance) and the number of compositional units $D_{\beta}(S')$ (which takes into account balance, variety and disparity) of industries in the US. The variety of industries, i.e. the number of different industry types that have at least one employee, is slightly increasing and then decreasing after 1950, with values ranging between 120 and 140 throughout the whole period. The sudden dip in variety in $1940$ is unexplained, and most likely due to data inconsistencies. The effective number of industries $D(S')$ shows a more pronounced hump-like pattern, with a period of diversification in $1850-1960$ in which the effective number of industries grows from 10 to around 80 industries, followed by a period of reconcentration. 

In contrast, the number of compositional units $D_{\beta}(S')$ shows a pattern of steady decline since $1900$, with values ranging from $10$ to $4$ compositional units. Hence, according to this measure the industrial diversity has decreased over the past centrury due to the fact that industries have become increasingly similar in terms of the occupations they employ. 

Considering the different diversities shown in Figure \ref{fig:IPUMS_divs}, it becomes clear that taking into account the different dimensions of diversity can lead to very different representations of the same data. In particular, considering only the variety may lead to an overestimation of diversity, as a the distribution over industries may still be concentrated in a few industries. Furthermore, if industries with similar occupational distributions are considered to be the same, the effective number of industries is still an overestimation of the diversity as some industries may be almost identical in terms of the occupations they employ. 

The effect of taking into account disparity of course depends on which features are considered. That, is, diversity may have increased when some other feature of industries is taken into account instead of their occupational distribution. Hence, application of these measures must be driven by the research question at hand. The interesting aspect of these measures however is that each dimension of diversity may show different dynamic, which is not visible when considering a single diversity index. Moreover, each dimension of diversity may play a distinct role in the system considered. Therefore, we consider variety, balance and disparity separately in the next section. 

\section{The 'ABC' decomposition}\label{sec:ABC} 
In order investigate the role of variety, balance, and disparity in practice, separate measures are required for each. To this end, I introduce the 'ABC decomposition', which decomposes diversity into its separate dimensions. Since $D_{\beta}(S')$ is a measure of diversity incorporating all three dimensions, a multiplicative decomposition into the variety (A), balance (B) and disparity (C) may be obtained as:
\begin{align}\label{eq:ABC}
D_{\beta}(S') = D_A(S') \cdot D_B(S') \cdot D_C(S').
\end{align} 

The variety $D_A$ is given by a simple count of the number of types in $S'$, or equivalently by the Hill number of order $q=0$ (see Appendix \ref{sec:app_hill}). The balance $D_B$ is computed by dividing the effective number of types in $S'$ (which takes into account both balance and variety) by the variety, leading to \cite{Hill1973} 
\begin{align*}
D_B(S') = \frac{D(S')}{D_A(S')} = \frac{D(S')}{n}.
\end{align*} 
$D_B(S')$ measures the evenness in the distribution of relative frequencies of the types. It takes values in $(0,1)$, with a maximum of $1$ that is attained when all relative frequencies are equal, i.e. $p_i = \frac{1}{n}$ for all types $i$ in $S$.
 
Since $D_{\beta(S'}$, $D_A(S')$ and $D(B')$ are then determined, the disparity $D_C(S')$ can be obtained by dividing the number of compositional units $D_{\beta}(S')$ (which takes into account all three dimensions) by the effective number as
\begin{align*}
D_C(S') = \frac{D_{\beta}(S')}{D(S')} = e^{-H(Y|X)}.
\end{align*} 
$D_C(S')$ can be considered as the diversity normalized for variety and balance, leaving a measure of disparity. It takes values in $(0,1)$, attaining the maximum value when none of the types have overlap in their features. The minimum is attained when all types have identical features. 

It is easily verified that \eqref{eq:ABC} holds with these definitions of $D_A(S')$, $D_B(S')$ and $D_C(S')$. The decomposition allows to study the three dimensions of diversity separately. The diversity $D_{\beta}(S')$ can be seen as the variety $D_A(S')$, corrected by the factors $D_B(S')$ and $D_C(S')$ which are both between $0$ and $1$. The variety can in turn be normalized by the total number of types in the classification considered to make it have values between $0$ and $1$ so that it is comparable to the balance and the disparity as a fraction of its maximum value. 

Applying the ABC decomposition to the example in Figure \ref{fig:example} leads to the results given in Table \ref{fig:table}. The results show, as expected, a decreasing disparity as the overlap between words increases. The decrease in disparity as the total number of letters decreases is accurately captured by the proposed measure. 

\begin{figure}[h]
\centering
\begin{tabular}{| l | c | c | c | c | c |}
    \hline
	& eff. number & ($\beta$-)diversity & variety & balance & disparity \\ 
	& $D(S')$ & $D_{\beta}(S')$ & $D_A(S')$ & $D_B(S')$ & $D_C(S')$ \\ \hline
	$\bf{A}$ & 3& $3$ & $3$ & $1$ & $1$  \\ 	 \hline
	$\bf{B}$ & 3& $2.08$ & $3$ & $1$ & $0.69$  \\ 	 \hline
	$\bf{C}$  & 3& $1.89$ & $3$ & $1$ & $0.63$ \\	 \hline
\end{tabular}

\caption{Values for the effective number of industries, the number of compositional units and the variety, balance and disparity as given by the ABC decomposition for the three examples depicted in Figure \ref{fig:example}. Variety and balance are equal in all three examples, as every industry employs three different occupations with equal weight. The disparity differs in all three cases, and is maximal for {\bf A}, in which there is no overlap of occupations between industries. For {\bf B} and {\bf C}, the measures show a lower disparity and hence a lower diversity for {\bf C}, in which the industries are composed of less occupations. }
\label{fig:table}
\end{figure} 

\begin{figure}[h]
\centering
\includegraphics[width=.8\textwidth]{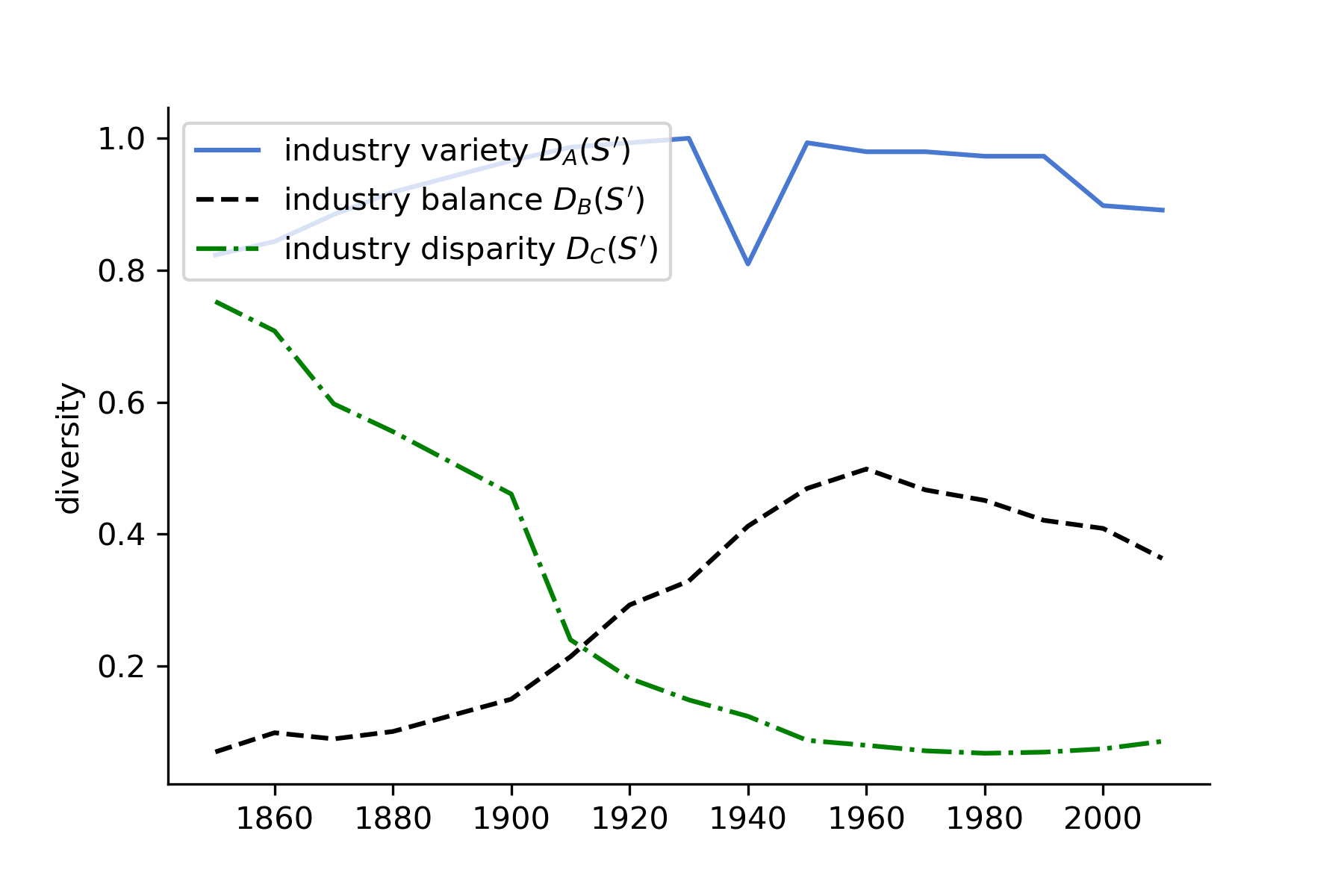} 
\caption{Variety, balance and disparity for industries given their occupational distribution. Variety is normalized by the total number of possible industries in the classification, which equals $147$. The variety of industries remains approximately constant over the whole period, and the balance shows diversification of industries up to $1960$, followed by a period of reconcentration. The disparity shows a decline over nearly the whole period, with a slight increase since $1980$.  }
\label{fig:IPUMS_ABC}
\end{figure}

Figure \ref{fig:IPUMS_ABC} shows the $ABC$ decomposition applied to the empirical example of industries in the US. It contains the same information as Figure \ref{fig:IPUMS_divs}, but shows the dynamics of variety, balance and disparity separately. Since all dimensions of diversity may move independently from each other, the ABC decomposition can help in analyzing the specific role of each dimension in different systems. 

\section{Multivariate extensions} \label{sec:infotheory} 
 From the framework of Hill numbers, an interesting relation follows between diversity and the information-theoretic notion of uncertainty. In particular, the beta diversity is given by 
\begin{align}\label{eq:Db}
D_{\beta}(S') &= \frac{D_{\gamma}(S)}{D_{\alpha}(S)} = e^{H(X) - H(X|Y)} = e^{MI(X,Y)}, 
\end{align}
where $MI(X,Y)$ denotes the mutual information between the random variables $X$ and $Y$\footnote{The mutual information is a measure of dependence between two random variables $X$ and $Y$, given by $MI(XY) = \sum_{ij} p_{ij} \log\left(\frac{p_{ij}}{p_ip_j}\right)$. It is nonnegative and symmetric, and can be interpreted as the average reduction in uncertainty about the outcome of one random variable, given knowledge about the outcome of the other.}. Taking the exponential of the mutual information between random variables $X$ and $Y$ translates it into a measure of diversity of the corresponding collection $S'$, discounted for the overlap in features given by $S$. Furthermore, the additive decomposition of information-theoretic measures corresponds to a multiplicative decomposition of diversities.  

\paragraph{Taking into account multiple features} 
Here, we generalize the diversity $D_{\beta}$ to take into account multiple features by exploiting the additive relations between multivariate information measures. For instance, returning to the example described in Figure \ref{fig:example}, we could add a feature to each word by color coding each letter, so that each word can be distinguished along two dimensions: its colors and its letters. Describing letters with random variable $X$, colors with random variable $Y$ and words with random variable $Z$, one can consider the joint probabilities $p_{ijk} = P(X=i, Y=j, Z=k)$ that a randomly sampled element is letter $i$, has color $j$ and is used in word $k$. In a network representation, the joint probabilities $p_{ijk}$ can be considered as the relative frequencies of hyperlinks between nodes $i$, $j$ and $k$ in a hypergraph that connects colors, letters and words to each other.

Following equation \eqref{eq:Db}, the diversity of words given their overlap in letters and colors is then given by 
\begin{align*} 
D^{XY}_{\beta}(S') = e^{H(XY) - H(XY|Z)}= e^{MI(XY,Z)},
\end{align*} 
where $H(XY) = -\sum_{ij}p_{ij} \log(p_{ij})$ is the Shannon entropy of the joint distribution $p_{ij}$, and the superscript in $D^{XY}(S')$ is used to indicate that diversity is taken with respect to the overlap in feature \textit{pairs} given by $XY$. Hence, every color-letter pair is interpreted as a distinct feature of a word. 

The effect of taking into account an additional feature on the diversity depends on the information contained by these features. In the current example, taking into account color as a second feature will not affect diversity much if colors and letters are highly correlated. On the other hand, diversity of words may be very large when colors and letters are independent of each other, thus capturing complementary information. Words that consist of the same letters may contain very different colors, and still add to the overall diversity. Mathematically, this can be seen by re-writing the beta diversity as (see Appendix \ref{sec:app2})
\begin{align*}
D^{XY}_{\beta}(S')  =e^{MI(X,Z) + MI(Y,Z)-MI(X,Y)+MI(XY|Z)},
\end{align*}
from which it is clear that diversity decreases when the dependence between features, given by  $MI(X,Y)$, increases. In the extreme case that letters $i$ and colors $j$ are independent, we have $MI(X,Y)=0$ so that (see Appendix \ref{sec:app2})
\begin{align*}
D^{XY}_{\beta}(S')  =e^{MI(X,Z) + MI(Y,Z)}= D_{\beta}^{X}(S')D_{\beta}^Y(S'),
\end{align*}
where $D_{\beta}^X(S')$ and $D_{\beta}^Y(S')$ denote the diversity of words with respect to the features described by random variables $X$ and $Y$, respectively. Thus, when features are independent the diversities that takes into account feature pairs can be obatined through multiplication of diversities that take into account the features separately. 

Results like this may be useful and relevant when estimating diversity from high-dimensional datasets containing multiple features. For example, one could consider the diversity of industries by not only taking into account the occupation but also the educational profile of people employed by an industry as a distinguishing characteristic. If educational profiles and occupations are uncorrelated, this diversity equals the product of the diversites that take into account occupations and educational profiles separately.

\paragraph{Aggregation} 
Another interesting interpretation is to consider the types in $S'$ to be an aggregation of the features $S$. In this setting the words are thus considered to be a specific way of aggregating letters. These words can in turn be further aggregated into sentences, effectively 'adding a layer' on top of the bipartite network depicted in Figure \ref{fig:example}. In such a setting, it follows from the current framework that the diversity of sentences depends only on the composition of words, and not on the composition of letters. The key assumption is that the two steps of aggregation are independent of each other, i.e. how words are aggregated into sentences is independent of how letter are aggregated into words. 

In the situation described above, the links between letters and words are given by the joint distribution $p_{ij}$ and the links between words and sentences by $p_{jk}$, where $k$ is the index for sentences, $i$ is the index for letters and $j$ the index for words. The probability of a letter-word-sentence triplet is then given by $p_{ijk} = p_{ij}p_{k|j}$. In other words, sentences and letters are conditionally independent on knowing the word, and their joint probability is given by 
\begin{align*} 
p_{ik} = \sum_j p_{i|j}p_{k|j}p_j,
\end{align*}
which implies that $MI(X,Z|Y)=0$. 
The diversity of sentences given the overlap in words and letters is then equal to the diversity when considering the overlap in words only (see Appendix \ref{sec:app3}):
\begin{align}\label{eq:agg}
D_{\beta}^Z(S') &= e^{MI(XY,Z)} = e^{MI(Z,Y)}. 
\end{align} 
Hence, when the composition of types described by $Z$ in terms of features described by $Y$ is independent of how the features $Y$ themselves are composed of other features $X$, the features $X$ are irrelevant for considering diversity of $Z$. 

As an example, consider the diversity of industries in Figure \ref{fig:IPUMS_divs}, where occupations are taken to be features of the industries. Hence, we can consider the distribution over industries as a particular way of aggregating over occupations. Similarly, occupations can be considered to be a collection of particular skills and tasks, and hence industries are, indirectly, also an aggregation of those skills and tasks. Equation \eqref{eq:agg} shows that as long as the composition of occupations in terms of skills and tasks is independent of the industry they are employed in, the diversity of industries is fully captured by occupations alone, and there is no need to consider skills and tasks. 

\section{Discussion}
This paper presented a framework to measure diversity while taking into account variety, balance and disparity of types. The framework builds on Hill numbers \cite{Hill1973, Jost2006} and the corresponding decomposition of diversity into independent alpha and beta components \cite{Jost2007}. 

The proposed framework provides a way to incorporate disparity into a measure of diversity without relying on an ad-hoc choice on pairwise similarities between types as in current approaches \cite{Daly2018, Leinster2012}. The framework also provides a natural derivation of diversities, solving the problem of how heavily each dimension of diversity should be weighed in a composite index \cite{Stirling2007}. Furthermore, the values of the proposed measure of diversity have a clear interpretation as the 'number of compositional units' \cite{Tuomisto2010}, describing the number of types that would be present in the hypothetical situtation in which there is no overlap in types and they have equal share.   

I have also proposed the 'ABC' decomposition of diversity that provides a way to capture variety, balance and disparity in separate measures. This measures disentangle the distinct dynamics and functional properties that different dimensions of diversity may have in different systems. In the context of economics for example, economic development is often associated with an increase of the diversity of economic activities \cite{Hidalgo2009, Saviotti1996}. It is however an open question what the role of the individual components of diversity is in the process of economic development - as the preliminary results in the current paper show, economic development may actually go hand in hand with decreasing disparity, when disparity is measured in terms of industries and the occupations they employ.

The proposed framework also reveals close connections between measuring diversity and information-theoretic measures of uncertainty. The simple additive properties of information-theoretic measures correspond to multiplicative properties of diversity measures, and enables derivation of simple properties when considering multiple features. These properties may provide useful tools in the analysis of high-dimensional datasets.

Furthermore, the diversity measure presented here can be interpreted as centrality measures on bipartite networks, or hypergraphs in the multivariate case. In this sense, the beta diversity captures strutural properties of the network. Application of these measure may be extended to directed networks (e.g. input-output tables in economics \cite{Leontief1966}), as any directed network may be interpreted as a bipartite network. 

In practice, it can be challenging to find the features of interest for a particular reasearch question. Such issues depend on the specific application at hand and the data available, and require theory-driven justification. A second challenge is to accurately estimate the proposed diversity measures from data. Estimation of information-theoretic measures like mutual information is notoriously difficult. Possible ways forward are to use bootstrapping techniques \cite{DeDeo2013} or Bayesian inference \cite{Hutter2001} for the estimation of diversity measures from data.

\section*{Acknowledgements}
I thank Andres Gomez-Lievano and Koen Frenken for their valuable feedback and comments. 

This work was funded by the Netherlands Organisation for Scientific Research (NWO) under the Vici scheme, number 453-14-014, and benefited from support from the Swaantje Mondt travel fund.

\bibliography{C:/GoogleDrive/PhD/library}{}

\newpage 

\appendix

\section{Hill numbers and entropy}\label{sec:app_hill}

In the definition of diversity we rely on the concept of Hill numbers, following \cite{Hill1973} and \cite{Jost2006}. The Hill number of order $q$ is given by the reciprocal of a generalized mean of the relative frequencies. The generalized weighted mean of the relative frequencies of types is given by 
\begin{align}\label{eq:mean} 
{^q}\bar{p} = \sqrt[q-1]{\sum_i p_i p_i^{q-1}},
\end{align} 
where the weights are given by the relative frequencies $p_i$. The parameter $q$ determines which mean is considered. For example, ${^0}\bar{p}$ denotes the Harmonic mean, ${^1}\bar{p}$ the geometric mean and for ${^2}\bar{p}$ the arithmetic mean \cite{Hill1973}. The Hill number of order $q$ measures the diversity of types as the reciprocal of the mean
\begin{align*} 
{^q} D(S) = \frac{1}{{^q}\bar{p}} = \left(\sum_i p_i^q \right)^{\frac{1}{1-q}}.
\end{align*} 

The parameter $q$ determines how heavily the average weights common or rare species. Values of $q>1$ weigh more heavily types with high relative frequency, and values of $q<1$ weigh more heaviliy the presence of types with smalle relative frequency. The minimal value of $q=0$ considers every type to contribute equally to the mean, regardless of its relative frequency. For $q=0$ the diversity is given by 
\begin{align*}
{^0}D(S) = \sum_i 1 = n
\end{align*} 
and gives simply a count of the number of types in $S$. The Hill number of order $0$ is thus a measure of \textit{variety}, which is also known as species richness in ecology. 

For $q=2$, one obtains
\begin{align*}
{^2}D(S) = \sum_i \frac{1}{p_i^2},
\end{align*} 
which relates directly to Simpson's index of concentration and the Gini-index  \cite{Jost2006}. \\
In general, the Hill numbers are related to the Rényi entropy \cite{Renyi1961} by ${^q}D(S) = e^{{^q}H(X)}$, where
\begin{align*}
{^q}H(X) = \frac{1}{1-q} \log \left(\sum_i p_i^q \right).
\end{align*} 
The Shannon entropy arises as a special case when taking the limit of $q \to 1$. This corresponds to the unique Hill number that does not favor either rare or common types and is given by
\begin{align*}
D(S) = \lim_{q\to 1}{^q}D(S) = e^{-\sum_i p_i \log(p_i)} = e^{H(X)}.
\end{align*} 
The relationship between Hill numbers and entropies described above tell us how to transform measures of uncertainty, given by entropies in units of bits or nats, into measures of diversity, given in units of the 'effective number of types'. The more uncertain one is about the type of a randomly sampled element from $S$ (i.e. the higher ${^q}H(X)$), the more diverse the set $S$ in considered to be. 

\section{Multiple features} \label{sec:app2}
\begin{figure}[h]
\centering
\includegraphics[width=\textwidth]{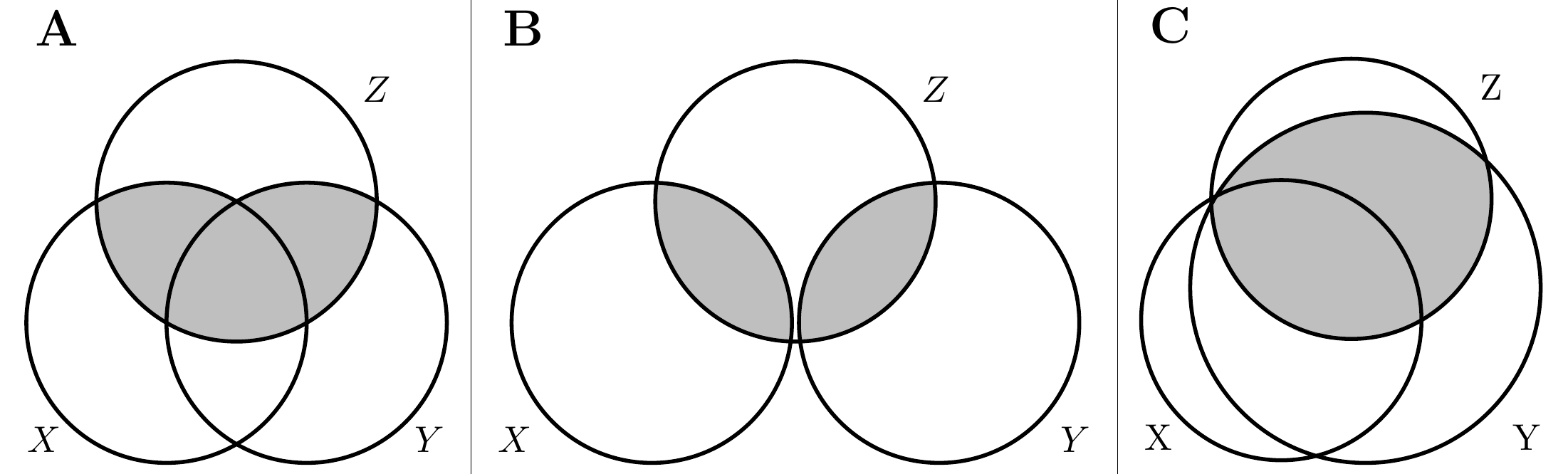} 

\caption{The entropies and mutual information can be represented using Venn diagrams, where each circle corresponds to the entropy $H(X)$ of the associated random variable $X$. The intersection of the two circles associated to $X$ and $Y$ for example, represents the mutual information $MI(X,Y)$, and their union represents the joint entropy $H(XY)$. The conditional entropy $H(X|Y)$ is given by subtracting the intersection from the total uncertainty $H(X)$. {\bf A} shows the mutual information $MI(XY,Z)$ from equation \eqref{eq:A}. The diversity of variable $Z$ given the overlap in features $XY$ is given by the exponential of the shaded area. {\bf B} shows the special case of \eqref{eq:B} in which the features $X$ and $Y$ are independent, i.e. $MI(X,Y)=0$. From the figure it is clear that $MI(XY,Z) = MI(X,Z) + MI(Y,Z)$, such that associated diversity in this case is multiplicative (equation \eqref{eq:B}). {\bf C} shows the case of \eqref{eq:C} in which $Z$ and $X$ are conditionally independent on $Y$, i.e. $MI(Z,X|Y)=0$. In this case, taking into account features $X$ becomes irrelevant in computing diversity of $Z$ given feature pairs $XY$. }
\label{fig:venn}
\end{figure}

This section elaborates on the results given in the main text on diversity when taking into account two features, described by random vriables $X$ and $Y$. The feature pairs are then described by the joint distribution $p_{ij} = P(X =i, Y = j)$. Using the simple additive properties of information-theoretic quantities, we show some simple results regarding diversities. The calculations are easily verified by considering the Venn diagrams in Figure \ref{fig:venn}.  

Here, we rewrite the diversity of types corresponding to random variable $Z$ given the overlap amond a pair of features given by random variables $X$ and $Y$ as 
\begin{align} \label{eq:A}
D^{XY}_{\beta}(S') &= e^{MI(XY,Z)}\\ \nonumber
&= e^{H(XY) - H(XY|Z)} \\ \nonumber
&= e^{H(X)+H(Y) - MI(XY) - H(X|Z) - H(Y|Z) +MI(XY|Z)} \\ \nonumber
&= e^{MI(X,Z) + MI(Y,Z) - MI(XY) + MI(XY|Z)},
\end{align} 
where we used that $H(XY) = H(X) + H(Y) - MI(XY)$ and $H(XY|Z) = H(X|Z) + H(Y|Z) - MI(XY|Z)$. From this, it becomes clear that the diversity becomes lower as the features $X$ and $Y$ have a larger dependence, i.e. are more correlated, as indicated by a large value of $MI(XY)$. 

In the special case that features $X$ and $Y$ share no information, i.e. $MI(X,Y)=0$, we have 
\begin{align} \label{eq:B}
D^{XY}_{\beta}(S') &= e^{MI(XY,Z)}\\ \nonumber
&= e^{MI(X,Z) + MI(Y,Z)}\\ \nonumber
&= D^{X}_{\beta}(S')D^{Y}_{\beta}(S').
\end{align} 
Hence, for independent features the diversities are multiplicative. 

\section{Aggregation} \label{sec:app3}
Another situation arises that we consider the types described by random variable $Z$ to be composed of features described by random variable $Y$, and the features $Y$ themselves have features described by a random variable $X$. Hence the links between types and features are given by the joint probability distribution $p_{jk}$, and the links between features and 'sub-features' by a joint distribution $p_{ij}$. When the joint probabilities $p_{ij}$ are independent of the joint probabilities $p_{jk}$, we have $p_{ijk} = p_{ij}p_{k|j} = p_{i|j}p_{k|j}p_j$. In other words, the random variables $Z$ and $X$ are conditionally independent given $Y$, which means that $MI(X,Z|Y)=0$. The diversity given feature pairs $XY$ can then be rewritten as
\begin{align} \label{eq:C} 
D_{\beta}^{XY}(S') &= MI(XY,Z)\\ \nonumber
&= e^{H(Z) - H(Z|XY)}\\  \nonumber
&= e^{H(Z) - (H(ZX|Y)- H(X|Y))}\\ \nonumber
&= e^{H(Z) - H(Z|Y)} = e^{MI(Z,Y)},
\end{align} 
where we used that $MI(X,Z|Y)=0$ implies that $H(XZ|Y) - H(X|Y) = H(Z|Y)$. In other words, considering $X$ is superfluous when considering the diversity of $Z$ in terms of features $XY$. 

\end{document}